\documentstyle[aps,twocolumn,epsf]{revtex}

\begin{document}

\draft

\title{
Strongly focused light beams interacting with single atoms in free space}

\author{S.J. van Enk and H.J. Kimble}

\address{Norman Bridge Laboratory of Physics, California
Institute of Technology 12-33, Pasadena, CA 91125}

\date{\today}
\maketitle

\begin{abstract}
We construct 3-D solutions of Maxwell's equations that describe Gaussian
light beams focused by a strong lens. We investigate the interaction of such
beams with single atoms in free space and the interplay between angular and
quantum properties of the scattered radiation. We compare the exact results
with those obtained with paraxial light beams and from a standard
input-output formalism. We put our results in the context of quantum
information processing with single atoms.
\end{abstract}
\section{Motivation}
The ability to manipulate small quantum systems individually is a necessary
requirement for quantum computing and quantum communication.  For example, in
order to perform single-qubit operations on a particular ion in an ion trap
quantum computer \cite{cz,wineland,steane} one has to focus a laser beam to
the position of that ion with a sufficiently high spatial resolution
\cite{nagerl}. Similarly, quantum communication protocols using atoms trapped
inside optical cavities \cite{cirac,enkQC} require single-bit and two-bit
operations on specific atoms. One question that arises is whether strong
focusing has an undesired side effect, namely, that the scattered light
contains information about the state of the qubit. The fear would be that the
laser intensity would have to be turned down so much, that the absence of a
photon from the laser beam becomes {\em in principle} detectable.

Conversely, if an atom in free space would indeed be able to modify
appreciably the state of a light field, then this effect could be used to our
advantage: a single atom could be used to perform quantum-logic operations on
single photons in free space \cite{sam}. An atom inside an optical cavity
strongly coupled to a cavity mode is known to be able to perform such tasks
\cite{turchette}, but such an experiment would be more straightforward to
conduct in free space.

At first sight the prospects of such an experiment seem good: the scattering
cross section of a two-level atom is $\sigma=3\lambda^2/(2\pi)$ for light of
wavelength $\lambda$ \cite{jackson}, and thus focusing light down to an area
$A< \sigma$ should be sufficient to induce a strong coupling. However, this
picture is too simplistic. Light beams are transversely polarized, which
implies that only part of the light entering the interaction region will
carry the polarization that the atom is sensitive to. Using a different
picture, since the atom would emit a dipole pattern (in a given $m\rightarrow
m'$ transition), it would be most efficiently excited by a field matching an
``incoming'' dipole field, as follows from time-reversal symmetry. Since a
focused laser beam does not have a large overlap with such a dipole pattern,
the effective absorption cross section is smaller than indicated by $\sigma$.

Early experiments \cite{peuse,itano} on the detection of single atom
fluorescence did not reach the strong focusing limit of $A\sim\lambda^2$.
Recently, however, impressive progress has been made in experiments on single
molecules in condensed matter and efficient detection of the fluorescence
light has become possible \cite{moerner,moerner2,fleury}. Experiments on
single atoms aiming at the strong focusing limit are underway as
well\cite{grangier}.

In a recent paper \cite{enkkim} we gave the results of explicit calculations
on the behavior of single atoms in free space irradiated by tightly focused
light beams. We were particularly interested in the quantum aspects of the
scattered light, and in evaluating how much a single atom is able to modify
the intensity and phase properties of the incident light. Here we present all
the details of that calculation and discuss possible extensions. These
results will be compared with similar calculations using (i) paraxial
Gaussian beams \cite{siegman} and (ii) a well-known quantum-optical
input-output formalism that was used in Refs.~\cite{car,car2} to study photon
statistics and intensity correlations of the field emitted by an atom in free
space. The latter model is basically a quasi one-dimensional model, with one
spatial variable describing the propagation of the light beam, and with one
additional parameter describing the solid angle subtended by the laser beam
at the atom's position (coinciding with the focal point of the light beam).
As we will demonstrate, however, neither this model nor, as expected,
paraxial beams accurately represent the case of a strongly focused light
beam.
\section{Strongly focused Gaussian beams}

Here we wish to calculate the field that one obtains by focusing a
monochromatic Gaussian (paraxial) beam by an ideal strong lens. We do that by
expanding the outgoing field (i.e. the field after the lens) in a complete
set of modes. In principle one can use any complete set of modes to describe
exact solutions of Maxwell's equations. In view of the cylindrical symmetry
of the problem we are interested in, the most convenient is to choose a set
that takes a simple form in cylindrical coordinates. In particular, we will
use a set of eigenmodes of 4 commuting operators corresponding to the
following 4 physical quantities: energy [with eigenvalue $\hbar
kc=\hbar\omega$ per photon \cite{note}], angular momentum in the $z$
direction [$m\hbar$], momentum in the $z$ direction [$\hbar k_z$], and
helicity [$s\hbar k_z/k$]. The modes are thus characterized by the 4 numbers
$\nu\equiv (k,k_z,m,s)$, which, once the field has been quantized, play the
role of quantum numbers. These modes were constructed in \cite{enk} to
clarify the meaning of orbital angular momentum of light \cite{orb}, and thus
by construction possess simple properties under rotations around the
propagation ($z$) direction.
 The complete
orthogonal set of modes $\vec{F}_{\nu}$  is defined such
that the free electric field (i.e. the solution of the
source-free Maxwell equations) can be expanded in this set
as
\begin{equation} \vec{E}=2 {\rm Re} [\sum_\nu a_\nu
\vec{F}_\nu \exp(-i\omega t))],
\end{equation}
with arbitrary complex amplitudes $a_\nu$. This requires the mode functions
to be transverse, i.e. $\nabla\cdot\vec{F}_{\nu}=0$. The summation over $\nu$
is a short-hand notation for
\begin{equation}
\sum_\nu\equiv \int{\rm d}k
 \int {\rm d}k_z \sum_s\sum_m.
\end{equation}
The dimensionless mode functions $\vec{F}_\nu$ in cylindrical coordinates
$(\rho,z,\phi)$ are defined by\cite{enk}
\begin{eqnarray}\label{modes}
\vec{F}_\nu(\rho,z,\phi)&=&\frac{1}{4\pi}
\frac{sk-k_z}{k}G(k,k_z,m+1)\hat{\epsilon}_-+\nonumber\\ &&+\frac{1}{4\pi}
\frac{sk+k_z}{k}G(k,k_z,m-1)\hat{\epsilon}_+\nonumber\\
&&-i\frac{\sqrt{2}}{4\pi}\frac{k_t}{k}G(k,k_z,m)\hat{z}
\end{eqnarray}
where $k_t=\sqrt{k^2-k_z^2}$ is the transverse part of the wave vector,
$\hat{\epsilon}_{\pm}=(\hat{x}\pm i \hat{y})/\sqrt{2}$ are the two circular
polarization vectors, and
\begin{equation}
G(k,k_z,m)=J_m(k_t\rho)\exp(ik_z z)\exp(im\phi),
\end{equation}
with $J_m$ the $m$-th order Bessel function. The mode functions satisfy the
orthogonality relations
\begin{eqnarray}
\int {\rm d} V \vec{F}^*_\nu(\vec{r})\cdot\vec{F}_{\nu'}(\vec{r})
=\delta(k-k')\delta(k_z-k_z')\delta_{mm'}\delta_{ss'}/k,
\end{eqnarray}
where the integration extends over all space. This follows directly from the
fact that the mode functions are eigenfunctions of commuting hermitian
operators.

For the remainder of this Section, we will consider only monochromatic beams
with a fixed value of $k=2\pi/\lambda$. For convenience we take $k_t$ as a
mode number instead of $k_z$ and we denote the reduced set of mode numbers by
$\mu\equiv (k_t,m,s)$, and introduce the notation
\begin{equation}
\sum_\mu\equiv \int {\rm d}k_t \sum_s\sum_m.
\end{equation}
For fixed $k$ the modes $\vec{F}_\mu$ are orthogonal in
planes $z=$constant:
\begin{eqnarray}\label{plane}
\int_{z={\rm constant}} {\rm d} S \vec{F}^*_\mu(\vec{r})\cdot
\vec{F}_{\mu'}(\vec{r})=\nonumber\\
\delta(k_t-k_t')\delta_{mm'}\delta_{ss'}/(2\pi k_t),
\end{eqnarray}
which is a useful relation for defining the action on light
beams of an ideal lens positioned in a plane $z$=constant.
\subsection{Focusing with an ideal lens}
The action of the lens is modeled here by assuming that the field
distribution of the incoming field is multiplied by a local phase factor
\begin{equation}\label{lens}
\varphi=\exp(-i k\rho^2/2f),
\end{equation}
with $f$ the focal length of the lens \cite{born,goodman}.  An ideal
parabolic lens would be represented by a phase factor
$\varphi_p=\exp(ik\sqrt{\rho^2+(f-\rho^2/2f)^2})$. In the paraxial limit $f
\gg \rho$ this factor becomes equivalent to $\varphi$. It may be that the
simple lens factor used here, $\varphi$, does not give rise to the strongest
possible focusing, and that $\varphi_p$ would improve on this. Moreover,
actual lens systems designed to focus light down to $A\sim \lambda^2$ consist
of {\em multiple} lenses (for instance, 12 in ongoing experiments
\cite{grangier}), partly to accommodate for the finite size, finite thickness
and other imperfections of real lenses as compared to ideal lenses. We
nevertheless  for convenience chose a single lens factor $\varphi$: it allows
for analytical evaluations and the class of light beams thus constructed does
reach the focusing limit of $A\approx \sigma$ (for instance, see the plots
corresponding to $f=100\lambda$). This is sufficient for our purposes of
showing that even focusing down to an area of the size of the atomic cross
section does not quite (by about a factor of $\sim 5$) lead to the strong
effects one may have expected or hoped for.

 If in the plane of the lens, say $z=0$, the
incoming beam is given by
\begin{equation}
\vec{F}_{{\rm in}}=\vec{F}_0(\rho,\phi),
\end{equation}
then the output field is given by
\begin{equation}\label{out}
\vec{F}_{{\rm out}}(\vec{r})=\sum_\mu \kappa_\mu\vec{F}_\mu(\vec{r}),
\end{equation}
with
\begin{equation}
\label{kappa} \kappa_\mu=2\pi k_t\int_{z=0} {\rm d}S \exp\left(-i
\frac{k\rho^2}{2f}\right)\vec{F}_0\cdot \vec{F}^*_\mu .
\end{equation}
This definition is such that the limit of $f\rightarrow\infty$ corresponds to
free-space propagation, as follows from the orthogonality relation
(\ref{plane}). Note that the field distribution (\ref{out}) is an exact
solution of Maxwell's equations, irrespective of the choice for $\vec{F}_0$
(in particular, we can take the incoming beam to be paraxial).

 If we approximate the incoming beam by a circularly
polarized (lowest-order) Gaussian beam with Rayleigh range $z_{{\rm in}}$
with $kz_{{\rm in}}\gg 1$ by\footnote{Here we assume for simplicity that the
focal plane of the incoming beam and the plane of the lens coincide.} its
dimensionless amplitude
\begin{equation}
\vec{F}_0(\rho,\phi)=\exp\left(-\frac{k\rho^2}{2z_{{\rm in}}}\right)
 \hat{\epsilon}_+,
\end{equation}
then $\kappa_\mu$ is given by
\begin{eqnarray}
\kappa_\mu&=&\delta_{m1}\pi k_t\frac{k_z+sk}{k} \int _0^\infty {\rm d}\rho
\rho J_0(k_t\rho) \times\nonumber\\&&\exp\left(-i\frac{k\rho^2}{2f}\right)
\exp\left(-\frac{k\rho^2}{2z_{{\rm in}}}\right).
\end{eqnarray}
This integral can be evaluated using
\begin{equation}
\int _0^\infty {\rm d}x x J_0(\beta x)\exp(-\alpha x^2)=
\frac{1}{2\alpha}\exp(-\beta^2/4\alpha),
\end{equation}
and gives the result
\begin{equation}
\kappa_\mu=\pi\delta_{m1}\frac{k_t}{k}\frac{k_z+sk}{k}\xi
\exp\left(-\frac{k_t^2}{2k}\xi\right),
\end{equation}
with
\begin{eqnarray}
\xi&=&z_R-iz_0\nonumber\\
 z_R&=&\frac{f^2z_{{\rm in}}}{z_{{\rm
in}}^2+f^2},\nonumber\\ z_0&=&\frac{fz^2_{{\rm in}}}{z_{{\rm in}}^2+f^2}.
\end{eqnarray}
The delta function $\delta_{m1}$ expresses the fact that a lens cannot absorb
angular momentum from a cylindrically symmetric light beam upon normal
incidence \cite{enk5,bavw}: the index $m$ of the outgoing beam is 1, because
the incoming beam has one unit of ``spin'' angular momentum.

When the paraxial limit is valid for the {\em outgoing} beam, i.e., when
$kz_R\gg 1$, $z_R$ and $z_0$ correspond, as we will show, to the Rayleigh
range and the position of the focal plane of the outgoing beam, respectively.
But also outside the paraxial limit, the focused light beam is characterized
by the two parameters $z_R$ and $z_0$.  The largest component of the output
field (\ref{out}) is the $\epsilon_+$ component, which is given by
\begin{eqnarray}\label{exact}
F_+&=&\frac{z_R-iz_0}{2} \int_0^k {\rm d}k_t \frac{k_t}{k}
\frac{2k^2-k_t^2}{k^2}\nonumber\\ &\times&
J_0(k_t\rho)\exp\left(-\frac{k_t^2}{2k}(z_R-iz_0)\right) \exp(ik_zz),
\end{eqnarray}
with $F_+\equiv\vec{F}\cdot\hat{\epsilon}^*$. Note here the similarity
between (\ref{exact}) and the expression given in \cite{allen} for a class of
light beams generalizing Laguerre-Gaussian (LG) beams \cite{siegman}. The
$\hat{z}$ and $\hat{\epsilon}_-$ components of the output field are
proportional to higher-order Bessel functions, $J_1(k_t\rho)$ and
$J_2(k_t\rho)$, respectively, and will therefore vanish on the $z$ axis. As
we will be interested in the interaction of an atom on axis with the focused
light beams, we will not consider these components here, but they may be
important in other circumstances.  Note as an aside that these two components
represent beams with 1 and 2 units of orbital angular momentum, and 0 and
$-1$ units of spin angular momentum, respectively, so that the total angular
momentum of the outgoing beam is indeed $\hbar$ per photon. See \cite{enkqo}
for a discussion how these different forms of angular momentum are
transferred to the internal and external angular momenta of an atom.

Returning to the
$\hat{\epsilon}_+$ component, when the paraxial approximation is valid for
the outgoing beam, we may take out a factor $\exp(ikz)$, use $k-k_z\approx
k_t^2/2k$, and extend the integration limit in (\ref{exact}) to infinity.
Defining
\begin{equation}
z_w=z_R+i(z-z_0),\end{equation}
these approximations lead to
\begin{eqnarray}
F_+&\approx&\frac{z_R-iz_0}{2}\exp(ikz) \int_0^\infty {\rm d}k_t
 2\frac{k_t}{k}\nonumber\\
&&\times J_0(k_t\rho)\exp\left(-\frac{k_t^2z_w}{2k}
\right)\nonumber\\ &=&\frac{(z_R-iz_0)\exp(ikz)}{z_w}\exp
\left(-\frac{k\rho^2}{2z_w}\right),
\end{eqnarray}
which, as announced, represents a Gaussian beam with
Rayleigh range $z_R$ and its focal plane located at
$z=z_0$. We can in fact rewrite the exact result
(\ref{exact}) into a different form that explicitly
displays the corrections to the paraxial approximation,
\begin{equation}
F_+=\exp(ikz)\frac{z_R-iz_0}{2} [F_1+F_2-F_3],
\end{equation}
with
\begin{eqnarray}
F_1&=&\left[\frac{2}{z_w}-\frac{2}{kz_w^2}
\left(1-\frac{k\rho^2}{2z_w}\right)\right]
\exp\left(\frac{-k\rho^2}{2z_w}\right),\nonumber\\
F_2&=&\int_0^k {\rm d}k_t
 \frac{k_t}{k}\frac{2k^2-k_t^2}{k^2}
 J_0(k_t\rho)\exp\left(-\frac{k_t^2z_w}{2k}\right)
\nonumber\\ &&\times
\left[\exp\left(i(k_z-k+\frac{k_t^2}{2k})z\right)-1\right],
 \nonumber\\
 F_3&=&\int_k^\infty {\rm d}k_t
 \frac{k_t}{k}\frac{2k^2-k_t^2}{k^2}
 J_0(k_t\rho)\exp\left(-\frac{k_t^2z_w}{2k}\right).
\end{eqnarray}
Outside the paraxial limit, when $kz_R$ is not large, the focal plane is no
longer at $z=z_0$ but moves towards the lens by several wavelengths, as shown
in Figures~1 and 2, where several examples of intensity profiles of focused
beams are plotted. Furthermore, unlike in the paraxial approximation, the
shape of the field is not just determined by the value of $z_R$, but depends
on $z_0$ as well.
\begin{figure}\label{f2} \leavevmode    \epsfxsize=8cm
\epsfbox{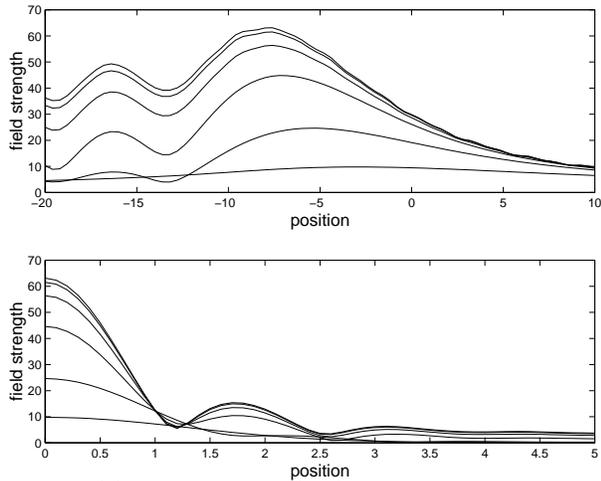}  \caption{(a) Field strength $|F_+|$ of strongly
focused Gaussian beams on the $z$ axis as a function of $Z\equiv
(z-z_0)/\lambda$. Note that the maximum field strength of the incoming field
$\vec{F}_0$ is 1. The lens is located at $z=0$ and is characterized by
$f=100\lambda$, so that $z_0\approx 100\lambda$. The incoming Gaussian beam
has increasing values of $z_{{\rm in}}/\lambda=1\times 10^3,3\times 10^3,
1\times 10^4\ldots 3\times 10^5$, resp., for the bottom to top curves. This
implies for the outgoing beam decreasing values of $z_R\approx
10\lambda,10\lambda/3,\lambda,\ldots \lambda/30$.
 (b) Field strength
in the focal plane as a function of the transverse
coordinate $\rho/\lambda$.
}\end{figure}
 The plots of the
transverse mode profiles show that beyond a certain point the width of the
field no longer decreases with stronger focusing. One cannot focus down a
laser field to below a certain limit, roughly about half a wavelength, with
the ideal lens with lens factor (\ref{lens}), no matter how small $z_R$
becomes. Moreover, one notes the asymmetry of the outgoing beam around the
focal plane, in contrast to a paraxial beam which is symmetric in its focal
plane. Since there is no a priori symmetry under reflections in the focal
plane, this fact should not be surprising.

\begin{figure}\label{f5} \leavevmode
\epsfxsize=8cm \epsfbox{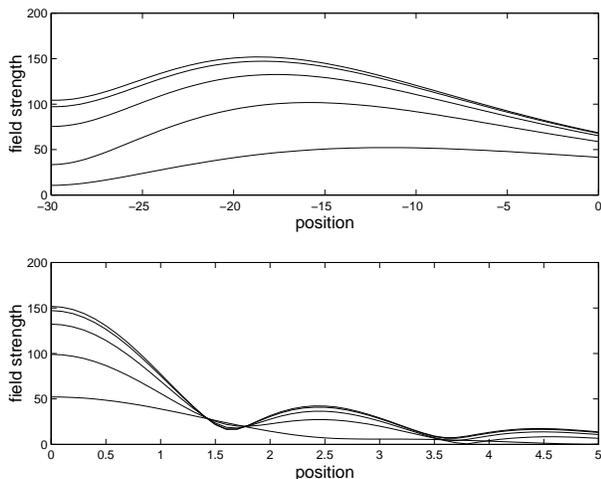} \caption{As Figure~1 but for
$f=500\lambda$ and $z_{{\rm in}}/\lambda=3\times 10^4, 1\times 10^5\ldots
1\times 10^6, 3\times 10^6$.}
\end{figure}
Let us note here that strongly focused light beams in general will display
phase singularities (rings in space where certain components of $\vec{E}$
vanish), of the kind investigated (both theoretically and experimentally) in
Refs.~\cite{marco,karman1,karman2,berry}. For Gaussian illumination of a
spherical lens, however, no such singularities appear \cite{karman1}.

Finally, it may also be interesting to consider tightly focused donut beams,
i.e., beams of light produced by focusing an incoming higher-order LG beam
\cite{siegman}. In the case of an incoming first-order LG beam, the incoming
field distribution $\vec{F}_0$ can be written as (again assuming its focal
plane coincides with the plane of the lens)
\begin{equation}\label{pm}
\vec{F}_0^{\pm}=\exp\left(-\frac{k\rho^2}{2z_{{\rm in}}}\right) \exp(\pm
i\phi) \frac{\rho}{z_{{\rm in}}} \frac{\hat{x}+i\hat{y}}{\sqrt{2}}.
\end{equation}
For the coefficients $\kappa^{\pm}_\mu$ we find then
\begin{eqnarray}
\kappa^{\pm}_\mu&=&\delta_{m 2,0}\pi k_t\frac{k_z+sk}{k} \int _0^\infty {\rm
d}\rho \frac{\rho^2}{z_{{\rm in}}} J_1(k_t\rho)\times\nonumber\\&&
\exp\left(-i\frac{k\rho^2}{2f}\right) \exp\left(-\frac{k\rho^2}{2z_{{\rm
in}}}\right),
\end{eqnarray}
where $\delta_{m 2,0}$ indicates that for the $+$ sign in (\ref{pm}) we get
$\delta_{m2}$ and for the $-$ sign $\delta_{m0}$. These delta functions again
express conservation of angular momentum: the outgoing beam possesses the
same angular momentum as the incoming beam \cite{enk5,bavw}, which has one
unit of ``spin'' angular momentum and $\pm 1$ units of ``orbital'' angular
momentum. The integral can be evaluated using
\begin{equation}
\int _0^\infty {\rm d}x x^2 J_1(\beta x)\exp(-\alpha x^2)=
\frac{\beta}{4\alpha^2}\exp(-\beta^2/4\alpha),
\end{equation}
and gives the result
\begin{equation}
\kappa_\mu=\pi\delta_{m 2,0}\frac{k_t^2}{k^2}\frac{k_z+sk}{k}
\frac{\xi^2}{z_{{\rm in}}} \exp\left(-\frac{k_t^2}{2k}\xi\right).
\end{equation}
For $\kappa^-_\mu\propto \delta_{m0}$ there is a nonzero field on axis of a
different polarization than the incoming field: the $z$ component, which
would be neglected in the paraxial limit is in fact the only nonvanishing
component on axis. In this case, however, the field on axis will be sensitive
both to deviations of the lens from an ideal spherical lens, and to
deviations of the incoming beam from a pure donut beam, even in the paraxial
limit. For some explicit examples of this sensitivity see, for instance,
Refs.~\cite{enk5,karman2}.

\section{Scattering light off of a single atom in free
space}\label{3} In this Section we will investigate the response of an atom
located in the focal region of a strongly focused laser beam of the form
(\ref{exact}) at $\vec{r}=\vec{r}_0$. We consider a $J_g=0\rightarrow J_e=1$
transition in the atom, as it is the simplest case where all three
polarization components of the light in principle play a role. For simplicity
we will assume the atom to be located on the $z$ axis, so that it in fact
interacts only with a single ($\hat{\epsilon}_+$) polarization component; as
mentioned above, the other two polarization components vanish on axis.

We are mostly interested in calculating the second-order
correlation function for the light field as a function of
position and time. For that purpose, the Heisenberg picture
is the most convenient. In the Heisenberg picture, the
electric field operator can be written as the sum of a
``free'' part and a ``source'' part \cite{vogel},
\begin{equation}
\vec{E}=\vec{E}_f+\vec{E}_s,
\end{equation}
where the free part is given by
\begin{eqnarray}
\vec{E}_f(\vec{r},t)&=&\sum_\nu \vec{F}_\nu(\vec{r}) a_\nu
 \exp(-i\omega t)+h.c.\nonumber\\
&\equiv& \vec{E}_f^{(+)}+\vec{E}_f^{(-)}.
\end{eqnarray}
Here we separated the field in positive- and negative-frequency parts and
used the spatial mode functions $\vec{F}_\nu(\vec{r})$ from
Eq.~(\ref{modes}), with $a_\nu$ the annihilation operator for mode $\nu$. The
source part for the case of a $J_g=0\rightarrow J_e=1$ transition is given by
\cite{vogel}
\begin{equation}\label{dip}
\vec{E}_s^{(+)}(\vec{r})=\sum_i\vec{\psi}_i(\vec{r}\prime)
\sigma_i^-(t-|\vec{r}\prime|/c),
\end{equation}
where $\vec{r}\prime=\vec{r}-\vec{r}_0$, and $\sigma_i^-$ is the atomic
lowering operator, and the sum is over three independent polarization
directions $i=\pm 1,0$. Equation (\ref{dip}) is valid in the far field, with
$\vec{\psi}_i(\vec{r})$ the dipole field
\begin{equation}
\vec{\psi}_i(\vec{r})=\frac{\omega_0^2}{4\pi\varepsilon_0c^2}
\left[\frac{\vec{d}_i}{r}- \frac{(\vec{d}_i\cdot\vec{r})\vec{r}}{r^3}\right].
\end{equation}
Here $\omega_0$ is the atomic resonance frequency, and $\vec{d}_i=d\hat{u}_i$
is the dipole moment between the ground state $|g\rangle$ and the excited
state $|e_i\rangle$ in terms of the standard unit circular vectors,
\begin{eqnarray}
\hat{u}_{-1}&=&\hat{\epsilon}_-,\nonumber\\
\hat{u}_0&=&\hat{z},\nonumber\\
\hat{u}_{1}&=&-\hat{\epsilon}_+,
\end{eqnarray}
and the reduced atomic dipole matrix element $d$.

Expressions containing the electric field in time-ordered and normal-ordered
form (as measured using standard photon detectors), such as the intensity and
the second-order intensity correlation function, can be transformed into what
Ref.~\cite{vogel} denotes as ${\cal O}$-ordered form, where $\vec{E}^{(+)}_s$
is placed to the left of $\vec{E}^{(+)}_f$, $\vec{E}^{(-)}_f$ to the left of
$\vec{E}^{(-)}_s$, and where the source parts are time-ordered. For instance,
if we assume the initial state of  the light field to be a coherent state
then the normally ordered intensity can be written as
\begin{eqnarray}\label{int}
I(t,\vec{r})&=&\langle\vec{E}^{(-)}(t,\vec{r})\cdot
\vec{E}^{(+)}(t,\vec{r})\rangle \nonumber\\
&=&\sum_{i,j=-1,0,1}\vec{\psi}_j^*(\vec{r})\cdot\vec{\psi}_i
(\vec{r})\sigma_{ee}^{ij}(t_r) +|\alpha|^2|\vec{F}_{{\rm
out}}(\vec{r})|^2\nonumber\\ &&+\sum_{i=-1,0,1}2{\rm
Re}[\alpha^*\exp(i\omega_0 t)\vec{F}_{{\rm
out}}^*(\vec{r})\cdot\vec{\psi}_i(\vec{r}) \sigma_{eg}^i(t_r)],\nonumber\\
\end{eqnarray}
where $\alpha$ determines the amplitude of the coherent state, such that
$\langle \vec{E}^{(+)}\rangle=\alpha \vec{F}_{{\rm out}}$. In Eq.~(\ref{int})
we introduced the retarded time $t_r=t-|\vec{r}^\prime|/c$, and
$\sigma_{eg}^i=\langle\sigma_i^-\rangle$ and
$\sigma_{ee}^{ij}=\langle\sigma_i^+\sigma_j^-\rangle$ are expectation values
of the corresponding atomic operators. The three terms in (\ref{int})
correspond to the intensity $I_d$ of the dipole field, $I_L$ of the incoming
laser beam, and the interference term. Similarly, the second-order
correlation function (where we now suppress the dependence of the fields on
$\vec{r}$)
\begin{eqnarray}
G^{(2)}(t,\tau,\vec{r})=\nonumber\\ \sum_{l,m=x,y,z}\langle E_l^{(-)}(t)
E_m^{(-)}(t+\tau) E_m^{(+)}(t+\tau)E_l^{(-)}(t)\rangle
\end{eqnarray}
consists of 16 terms. For $\tau=0$, 7 of those vanish identically, and the
remaining ones are
\begin{eqnarray}
G^{(2)}(t,0,\vec{r})=|\alpha|^4|\vec{F}_{{\rm out}}|^4
+\sum_{i,j}2|\alpha|^2|\vec{F}_{{\rm out}}|^2
\vec{\psi}^*_i\cdot\vec{\psi}_j\sigma_{ee}^{ij}(t_r)\nonumber\\ +\sum_{i}
4{\rm Re} (\alpha^*\exp(i\omega_0 t) )\vec{F}^*_{{\rm out}}\cdot\vec{\psi}_i
|\alpha|^2|\vec{F}_{{\rm out}}|^2\sigma^i_{eg}(t_r))\nonumber\\ +\sum_{i,j}
2|\alpha|^2(\vec{F}_{{\rm out}}\cdot \vec{\psi}^*_i) (\vec{F}_{{\rm
out}}^*\cdot\vec{\psi}_j\sigma_{ee}^{ij}(t_r)).
\end{eqnarray}
For the evaluation of the atomic quantities we assume the atom reaches a
steady state. Define $C_i= \alpha \vec{d}_i^*\cdot \vec{F}_{{\rm
out}}(0)/\hbar$ and the matrices ${\cal M}$ and ${\cal M}_{1,2}$ by
\begin{eqnarray}
{\cal M}_1^{ij}&=&\frac{C_iC_j^*/\Gamma}
{\Gamma/2+i\Delta}\nonumber\\
{\cal
M}_2^{ij}&=&\frac{C_iC_j^*/\Gamma}
{\Gamma/2-i\Delta}\nonumber\\ {\cal M}&=&({\cal M}_1+{\cal
M}_2)/({\cal M}_1+{\cal M}_2+\openone)
\end{eqnarray}
with $\Gamma$ the decay rate of the excited states
\cite{vogel},
\begin{equation}
\Gamma=\frac{d^2\omega_a^3}{3\pi\hbar\epsilon_0c^3}
\end{equation}
and $\Delta=\omega_0-\omega_a$ the detuning of the laser
field from atomic resonance. In the steady state we have
then
\begin{eqnarray}
\sigma_{gg}&=&\frac{1}{1+{\rm Tr} {\cal M}},\nonumber\\
\sigma_{ee}&=&\sigma_{gg}{\cal M},\nonumber\\
\sigma_{eg}&=&\frac{i\sigma_{gg}\vec{C}-i\sigma_{ee}
\vec{C}}{\Gamma/2-i\Delta}\exp(-i\omega_0 t).
\end{eqnarray}
We are mainly interested in finding the maximum effect the atom may have on
the outgoing beam.  We therefore consider the case of weak on-resonance
excitation, i.e., $|\vec{C}|\ll\Gamma$ and $\Delta=0$. We then calculate
$g^{(2)}(\tau,\vec{r})\equiv G^{(2)}(\tau,\vec{r})/I^2(\vec{r})$ at
$\tau=0$---there is no dependence on $t$ in the steady state and for
simplicity we leave the argument $t$ out--- as a function of position in the
far field. Results are shown in Figures~3--4. The distance to the atom is
fixed at $R=50\lambda$ for numerical reasons. Note that the angular spectrum
does depend on the precise value of $R$. Only in the forward direction do the
dipole field and a laser field display the same asymptotic behavior.
 In the forward direction, i.e. on the $z$ axis, the
laser field turns out to overwhelm the scattered field, irrespective of how
strongly the light is focused onto the atom. This may be compared to a
similar result for classical scattering from spherical dielectrics with light
focused down to spot sizes larger than 5 times the size of the spheres
\cite{hodges}. Hence we find that $g^{(2)}(0,\vec{r})\approx 1$ for forward
scattering, which is in sharp contrast with the result from \cite{car} which
predicts a large bunching effect (i.e., $g^{(2)}\gg 1$) for tight focusing.
The figure also shows that in a perpendicular direction the dipole field
dominates, so that $g^{(2)}(0,\vec{r})=0$ for $\phi\rightarrow\pi/2$ (i.e.,
there the light is almost purely fluorescence light, which is anti-bunched
\cite{carw,mandel}). $g^{(2)}$ reaches a maximum around angles where the
scattered and laser fields are comparable in magnitude. The oscillations
indicate that $g^{(2)}(0,\vec{r})$ is very sensitive to the relative phase
between the dipole and the laser field. In fact, maxima in $g^{(2)}$ appear
when the free field and the dipole field interfere destructively. Indeed,
this implies that the total field is smaller than the laser field, which
implies a photon has just been absorbed by the atom. The atom is therefore in
its excited state, and hence one can expect a fluorescent photon to appear
soon, thus leading to a strong bunching effect.

Going from Fig.~3 to 4 corresponds to tighter focusing
($z_R$ decreases by a factor of 2) and we see that
\begin{enumerate}
\item In the forward direction the ratio of the amounts of laser
and scattered light decreases (but it's still much larger
than 1).
\item The region where $g^{(2)}$ reaches its maximum moves
outward to larger angles $\phi$.
\item The ratio of the amounts of laser and scattered
light at $\phi=90^o$
increases by a large amount.
\end{enumerate}
\begin{figure}\label{fig3} \leavevmode    \epsfxsize=8cm
\epsfbox{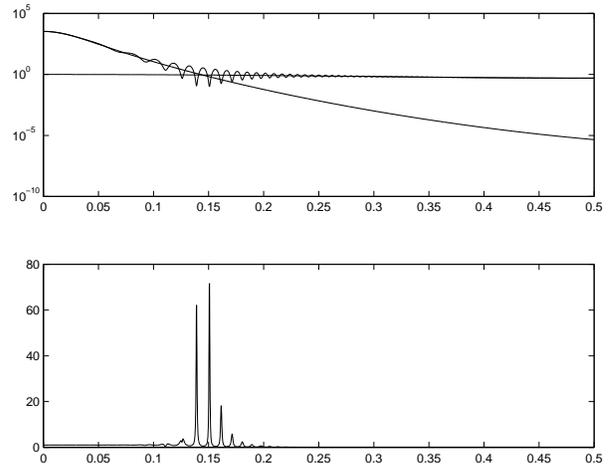}  \caption{ Plot (a) gives the relative intensities of
the laser field, the dipole field and the total field as a function of the
angle $\phi/\pi$ with the $z$ axis (i.e, at position
$\vec{r}=[R\sin\phi,0,R\cos\phi]$ where we chose $R=50\lambda$ here and for
all further calculations. The parameters for the incoming beam and the lens
are $f=500\lambda$ and $z_{{\rm in}}=3\times 10^4 \lambda$, so that
$z_R=8.3\lambda$ and $z_0=500\lambda$, where we chose $\lambda=852$nm,
corresponding to the D2 transition in Cs, and the atomic dipole moment $d$
adjusted so as to give the correct corresponding spontaneous emission rate
$\Gamma=2\pi \times 2.6$MHz for the 6$P_{3/2}$ states of Cs. Plot (b) gives
$g^{(2)} (0,\vec{r})$ as a function of $\phi/\pi$. }
\end{figure}
\begin{figure}\label{f5c} \leavevmode    \epsfxsize=8cm
\epsfbox{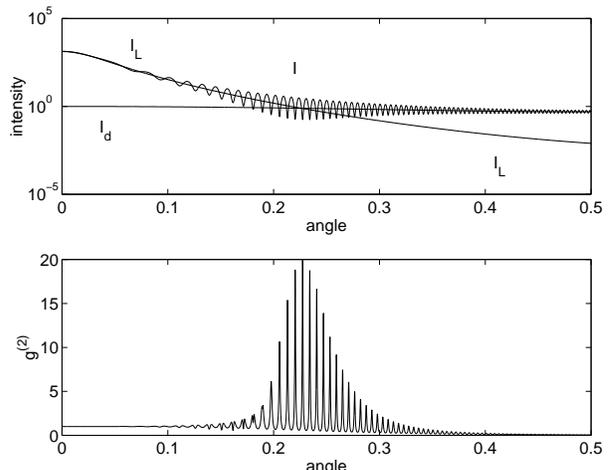}  \caption{ As Figure~\ref{fig3}, but the parameters
for the incoming beam and the lens are $f=500\lambda$ and $z_{{\rm
in}}=6\times 10^4 \lambda$, so that $z_R=4.2\lambda$ and $z_0=500\lambda$.}
\end{figure}
We can compare these results with those for a Gaussian beam
with the same beam parameters. Figures~5 and 6 show that
the 3 conclusions still hold. However,
 a Gaussian beam exaggerates
the amount of light in the forward direction (small $\phi$) at the cost of
greatly underestimating it for larger angles. This implies that the region
where $g^{(2)}$ reaches its maximum is moved to smaller angles $\phi$ for a
paraxial beam.
\begin{figure}\label{f5p} \leavevmode    \epsfxsize=8cm
\epsfbox{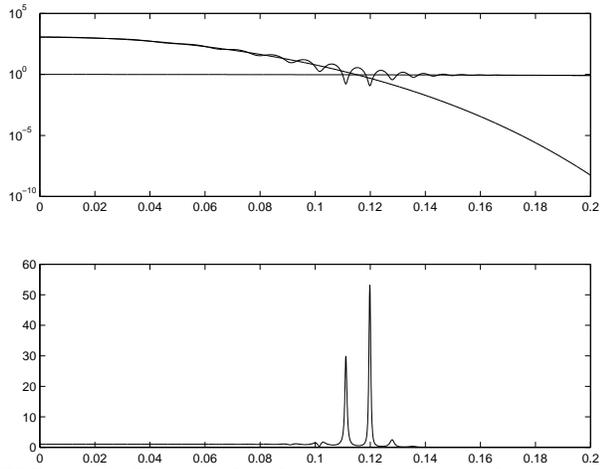}  \caption{ As Figure~3, but for a paraxial beam
characterized by the same beam parameters $z_R=8.3\lambda$ and
$z_0=500\lambda$.}
\end{figure}
\begin{figure}\label{f5p2} \leavevmode    \epsfxsize=8cm
\epsfbox{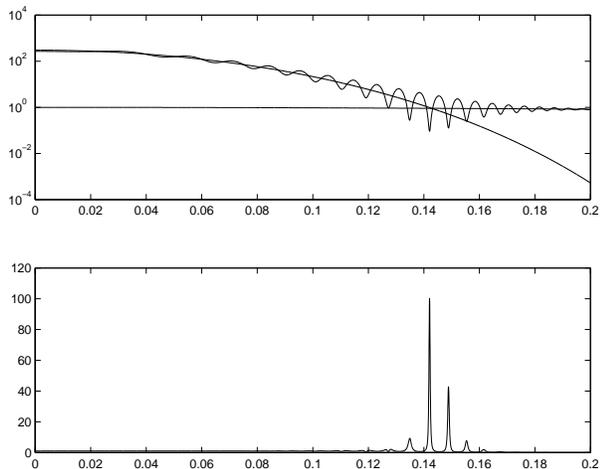}  \caption{ As Figure~4, but for a paraxial beam
characterized by the same beam parameters $z_R=4.2\lambda$ and
$z_0=500\lambda$.}
\end{figure}
We now focus on forward scattering, and plot in Figure~7 the ratio of the
intensities of the laser field and the dipole field, i.e.,
$K=|\vec{E}_f|^2/|\vec{E}_s|^2$, in the forward direction ($\phi=0$) as a
function of the normalized (dimensionless) beam width $w$, defined as
\begin{equation}
w=\sqrt{\frac{z_R}{\pi \lambda}}.\end{equation}
 The laser field intensity is
seen to be much larger than the dipole field intensity, by at least a factor
of $\sim 500$. For a Gaussian beam, on the other hand, the ratio becomes
arbitrarily small for small $w$. This has immediate consequences for the
value of  $g^{(2)}(0,\vec{r})$ (see Figure~8).
\begin{figure}\label{ratio} \leavevmode    \epsfxsize=8cm
\epsfbox{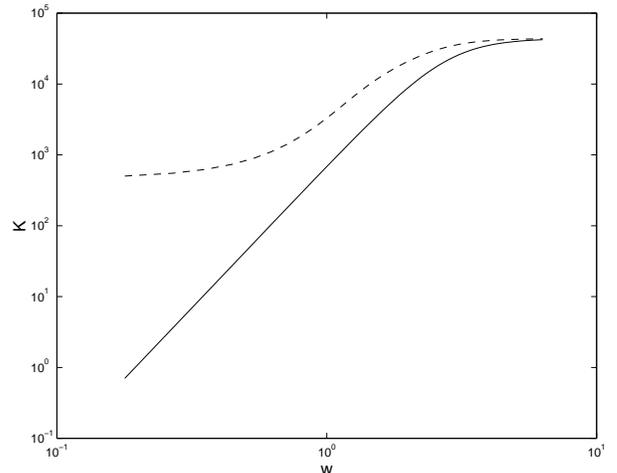}  \caption{ The relative intensity of the laser beam to
the dipole field in the forward direction, $K$, as a function of the beam
width parameter $w$ for the case $f=500\lambda$. Since $z_R\leq f/2$, the
beam width $w$ satisfies $w\leq 6.3$. The dashed curve corresponds to the
exact solution, the solid curve to a Gaussian beam.}\end{figure}
 For a
Gaussian beam, the intensity in the focal region is not bounded. In fact, for
decreasing values of $w$, more and more energy is concentrated in the focal
region, so much so that the dipole field will eventually dominate the field
in the forward direction. In that case, the forward direction will display
anti-bunching (for $w$ smaller than approximately 0.07). Before that,
however, $g^{(2)}$ reaches a large maximum at about $w\approx 0.2$, namely
when the dipole and laser fields are comparable in magnitude.  For larger
values of $w$, the scattered field will be negligible, and
$g^{(2)}\rightarrow 1$, but only after reaching a minimum close to zero
around $w=0.3$. The latter characteristics were also found in
Ref.~\cite{car}.
 For the exact solutions, however, none of these effects is
present, and the laser field always dominates the dipole
field, so that $g^{(2)}\approx 1$ for all values of the
beam width.
\begin{figure}\label{G2} \leavevmode    \epsfxsize=8cm
\epsfbox{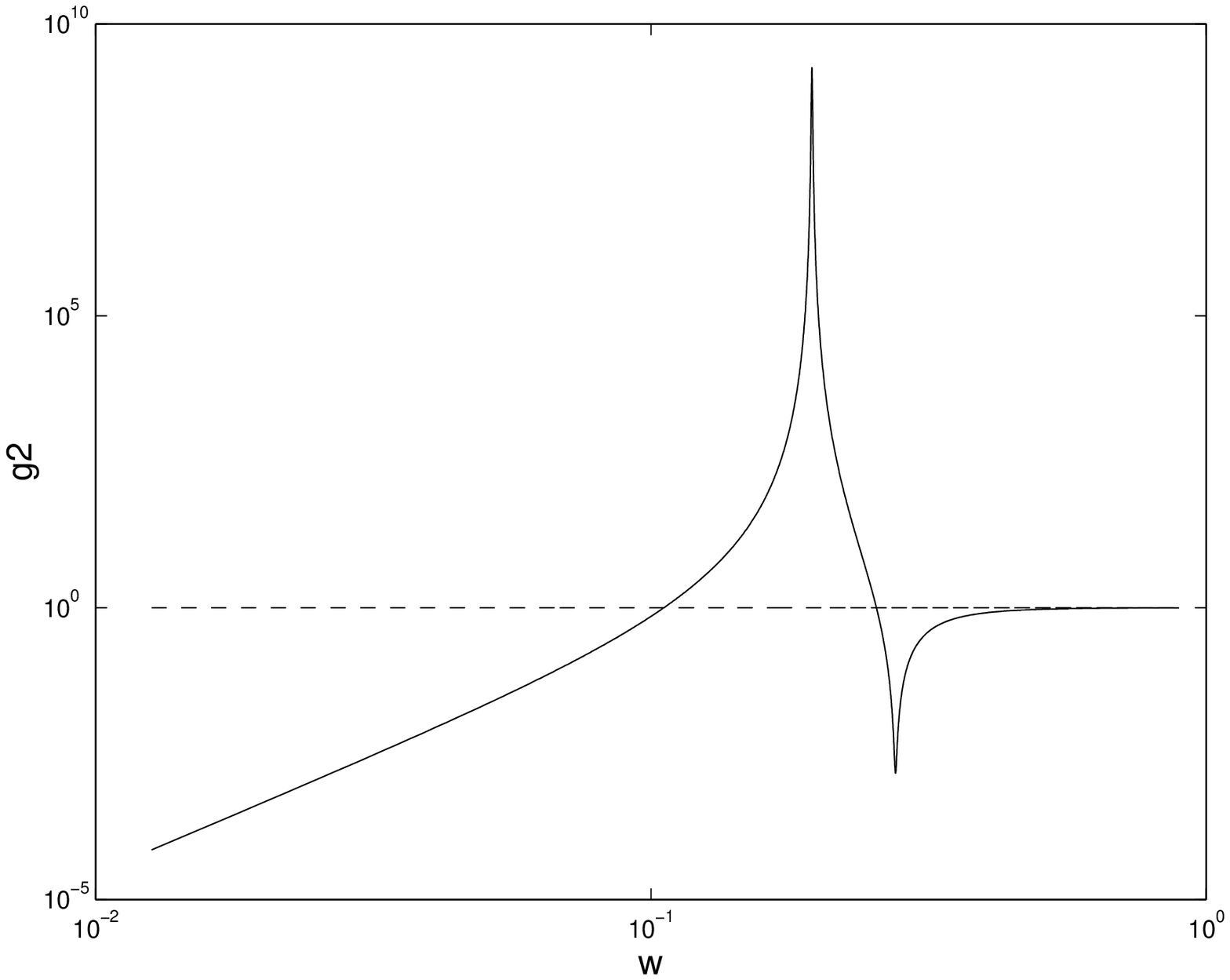}  \caption{ $g^{(2)}(0,\vec{r}))$ in the forward direction
as a function of the beam width parameter $w$. The solid curve corresponds to
a Gaussian beam, the dashed curve to the exact solution
\protect{(\ref{exact})}.}
\end{figure}
Let us finally quantify the effects of focused light on atoms in a different
way by considering the following. If the atomic dipole is $\vec{d}$, then the
relevant quantity determining the excitation probability of an atom is
 $|\vec{d}\cdot \vec{E}^{(-)}(\vec{r
}_{0})|^{2}$ evaluated at the atom's position $\vec{r}_{0}$, while the total
incoming energy flux is given by $\int {\rm d}S|\vec{E}^{(-)}|^{2}$. In
contrast to the naive expectation $ R\sim\sigma/A$, the actual ratio $R_{s}$
that determines the fraction of the energy incident on the atom which will be
scattered is given by
\begin{equation}
R_{s}=\frac{3\lambda ^{2}|\hat{d}\cdot \vec{E}^{(-)}(\vec{r}_{0})|^{2}}{2\pi
\int {\rm d}S|\vec{E}^{(-)}|^{2}}.
\end{equation}
This ratio is plotted in Fig.~9 as a function of the width $w=w_{R}/\lambda $
for several values of the focal parameter $f$. For smaller $f$ the best
achievable ratio increases, as expected, but for realistic lens parameters,
the optimum $R_{s}$ is about 20\%. Even for small values of $f$, the maximum
scattering ratio does not go beyond 1/2. This can be understood by noting
that the optimum shape of the illuminating field would be a dipole field.
Here with light coming only from one direction, one may expect $R_s$ to be at
most 1/2. Obviously, with one mirror behind the atom, one can improve the
scattering ratio $R_s$ by a factor of 2. And of course, by building an
optical cavity around the atom the atom-light interaction can be enhanced by
many orders of magnitude, but that's a different story \cite{cqed}.
\begin{figure}[tbp]
 \leavevmode    \epsfxsize=8cm
\epsfbox{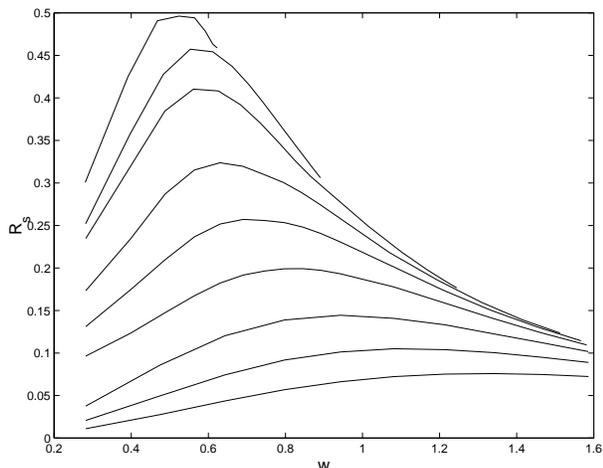} \caption{ The scattering ratio $R_{s}$ as a function
of the normalized beam
width $w=w_{R}/\protect\lambda $. The different curves correspond to $f/%
\protect\lambda =2.5,5,10,25,50,100,250,500,1000$ resp., for top to bottom
curves.}
\end{figure}

\section{Discussion and conclusions}
We constructed propagating wave solutions of Maxwell's equations describing
tightly focused laser beams. The method we used consisted of expanding the
outgoing beam in a complete set of solutions and matching it at the plane of
a lens to a given incoming beam. The lens was assumed ideal (infinitely thin)
and the incoming beam was chosen to be Gaussian.

 We then investigated quantum-statistical properties of the light
emitted by an atom in free space, when it is illuminated by such a beam.
Light detected in the forward direction does not display any bunching, nor
anti-bunching effects: the field is dominated by the laser light, and the
normalized second-order intensity correlation function is practically unity.
This may not be surprising but is in contrast to results obtained by using
Gaussian beams  and by a standard quantum-optical input-output model.
Gaussian beams are no longer valid approximate solutions under strong
focusing conditions, and in particular exaggerate the focal intensity by a
large amount.  On the other hand, the input-output formalism implicitly
assumes that the scattered field propagates in the same manner as the
incident light beam: in free space this would correspond to illumination with
a laser field whose profile mimics the dipole pattern. Inside a cavity,
however, the model is expected to apply, as the situation there is to a good
approximation one-dimensional. Indeed, the equations ultimately  assume the
same form as those for an atom coupled to a cavity mode in the bad-cavity
limit \cite{rice}.

Although the model of a single ideal lens with a simple lens factor of
Eq.~(\ref{lens}) may not lead to the strongest possible focusing
\cite{grangier}, the amount of focusing reached is sufficiently strong
(focusing areas $A$ less than or equal to the absorption cross section
$\sigma=3\lambda^2/(2\pi)$ of an atom) to conclude that the interaction a
focused light beam with an atom is not as strong as might be expected on the
basis of the ratio $\sigma/A$ (see Fig.~9). Its consequence for quantum
information processing may be phrased as: in free space it is easier and more
efficient to use light to process quantum information carried by an atom,
than to use an atom to process quantum information carried by photons.

\section*{Acknowledgments}
SJvE thanks Ph.~Grangier for illuminating discussions on focusing light
beams. We thank R.~Legere for helpful comments and discussions.  This work
was funded by DARPA through the QUIC (Quantum Information and Computing)
program administered by the US Army Research Office, the National Science
Foundation, and the Office of Naval Research.


\begin{references}
\bibitem{note}Of course, at this point the
electromagnetic field is just a classical solution of the Maxwell equations
and reference to photons and $\hbar$ is not necessary: we could have defined
energy density, momentum density etc.\ instead.
\end{references}
\end{document}